\documentclass[%
 aip,
rsi,%
 amsmath,amssymb,
 reprint,%
]{revtex4-1}

\usepackage{color}
\usepackage{graphicx}
\usepackage{dcolumn}
\usepackage{bm}
  \usepackage{algorithm} 
  \usepackage{algpseudocode}  
  \usepackage{multirow}
  \usepackage{natbib}
\usepackage{tikz}
\usetikzlibrary{shapes.geometric, arrows}

\begin{document}

\preprint{AIP/123-QED}

\title{ 
Efficient Computation of Free Energy Surfaces of Chemical Reactions using Ab Initio Molecular Dynamics with Hybrid Functionals and Plane Waves
%
%
}

\author{Sagarmoy Mandal}
\affiliation{ 
Department of Chemistry, Indian Institute of Technology Kanpur, Kanpur - 208016, India
}%

\author{Nisanth N. Nair}
\email{nnair@iitk.ac.in}

\affiliation{ 
Department of Chemistry, Indian Institute of Technology Kanpur, Kanpur - 208016, India
}%

\date{\today}

\begin{abstract}
{\em Ab initio} molecular dynamics  (AIMD) simulations employing density functional theory (DFT) and plane waves
are routinely carried out using density functionals at the level of Generalized Gradient Approximation (GGA).
AIMD simulations employing hybrid density functionals are
of great interest as it offers more accurate description of  
structural and dynamic properties than the GGA functionals.
However, computational cost for carrying out calculations using hybrid functionals and plane wave basis set is
at least two order of magnitude higher than  GGA functionals.
Recently, we proposed a strategy that combined the adaptively compressed exchange operator formulation and the multiple time step integration scheme to reduce the computational cost 
about an order of magnitude [{\em  J. Chem. Phys.} {\bf 151}, 151102 (2019)]. 
In this work, we demonstrate the application of this method to study chemical reactions, in particular, formamide hydrolysis in alkaline aqueous medium.
By actuating our implementation with the well-sliced metadynamics scheme,
we are able to compute the two-dimensional free energy surface of this reaction at the level 
of hybrid-DFT. 
Accuracy of PBE0 (hybrid) and PBE (GGA) functionals
in predicting the free energetics of the chemical reaction is investigated here.
%
\end{abstract}

\maketitle

\section{\label{sec:intro}Introduction}
{\it Ab initio } molecular dynamics (AIMD) simulations, where the atomic forces are computed
on the fly using first-principles based quantum mechanical methods, is widely used to 
investigate chemical reactions.\cite{marx-hutter-book,tuckerman_aimd,Tuckerman_2002_aimd,Payne_RevModPhys}
%
%
%
%
%
Specifically, for computing free energetics and mechanism of chemical reactions in liquids and heterogeneous interfaces, AIMD technique is preferred over the static quantum chemical
approaches that rely on finding stationary points on the potential energy landscape.
The combination of Kohn Sham density functional theory (KS-DFT) and plane wave (PW) basis sets is the method of choice for carrying out AIMD simulations of periodic and isolated systems.\cite{marx-hutter-book}
%
%
%
%
This can be attributed to the fact that PW basis set is free from basis set superposition errors and Pulay forces, and is inherently 
periodic.
%
%
%
%
%
%

The quality of the results obtained from a KS-DFT calculation critically depends on the chosen exchange-correlation (XC) functional.
For the past two decades or more, AIMD using PW KS-DFT is largely restricted to the Generalized Gradient Approximation (GGA)\cite{PRA_GGA_Becke,PRB_GGA_LYP,PRL_GGA_PBE} level of XC functionals.
%
%
However, it is well known that these functionals suffer from self interaction error (SIE).\cite{Chemist's_Guide,Science_DFT_limitations,PRB_SIC,PRL_DFT_errors}
%
%
%
%
SIE leads to over-delocalization of electron density,
%
resulting in the underestimation of chemical reaction barriers, band gap of solids, and dissociation energy.\cite{Science_DFT_limitations,PRL_DFT_errors} 
Also, SIE severely affects the electronic structure properties of open-shell systems.

%
%
In a hybrid-type XC functional, a portion of the Hartree-Fock exchange energy is added to the GGA exchange energy.\cite{Chemist's_Guide,Martin-book,JCP_B3LYP,JCP_PBE0,JCP_HSE}
KS-DFT calculations with hybrid functionals are generally known to improve the prediction of energetics, structures, electronic properties, chemical reaction barriers and band gap of solids.\cite{JCP_B3LYP,JCP_PBE0,JCP_PBE0_model,JCP_HSE,PCCP_HFX,PCCP_HSE,Galli_RSB_JPCL,Chem_Rev_Cohen} 
Also, hybrid functional based AIMD simulations give an improved description of the structural and dynamical properties of liquids.\cite{JPCB_AIMD_HFX,JCTC_AIMD_HFX,JCP_AIMD_HFX,Mol_Phy_Car_MLWF,Mol_Phy_Car_MLWF_1,JPCB_water_hfx} 
Hybrid functionals reduce SIE and improve the accuracy of KS-DFT computations.
Recently, hybrid functional based AIMD simulations in combination with enhanced sampling methods are shown to improve the accuracy of the computed free energy surfaces.\cite{JCTC_Galli_FES,JCP_sagar}
%
%
%
However, AIMD with hybrid functionals and PW basis set 
has a huge computational overhead, which is associated with the evaluation of the
exact exchange energy.\cite{JCP_HFX_Voth} 
%
%
%
%
Thus, free energy calculations of chemical reactions, that requires $\sim10^6-10^7$ force evaluations, in systems containing several hundred or more atoms, are seldom performed employing AIMD.

One of the ways to decrease the computational time is by evaluating the exchange integral using localized KS orbitals.\cite{PRB_Car_Wannier,JCP_AIMD_HFX,Mol_Phy_Car_MLWF,Mol_Phy_Car_MLWF_1,Nature_Car_MLWF,PRL_RSB,JCTC_RSB,JCTC_RSB_1,Galli_RSB_CPL,Galli_RSB_JPCL,Galli_RSB_JPCL1,Car_hfx_2019,JCP_sagar,JCC_HFX_Jonsson,HFX_Curioni,Espresso_hfx,Carnimeo_2019}
Usage of the multiple time step (MTS) algorithms,\cite{MTS_1,MTS_2,MTS_3,MTS_4,r-RESPA}
in particular the reversible reference system propagator algorithm (r-RESPA),\cite{r-RESPA}  
is a promising alternative.
%
%
%
In the r-RESPA algorithm, computationally cheap fast forces are computed more frequently as compared to the computationally costly slow forces.
%
%
%
In this way, one obtains the required speed-up using r-RESPA in AIMD simulations.\cite{MTS_AIMD_Carter,MTS_AIMD_PRE,HFX_Hutter_JCP,MTS_AIMD_Steele_1,MTS_AIMD_JCP,MTS_AIMD_Steele_2,MTS_AIMD_Steele_3,MTS_AIMD_Ursula}
%
%
%
%
In order to apply r-RESPA method in hybrid functional based AIMD simulations, an artificial time scale separation in ionic forces has to be created.
In some of the earlier works, time scale separation was introduced by combining the forces from GGA and hybrid functionals\cite{HFX_Hutter_JCP,MTS_AIMD_Ursula}
or forces from different levels of two-electron integral screening.\cite{MTS_AIMD_Steele_3}
Recently, we proposed a r-RESPA scheme based on the 
adaptively compressed exchange (ACE)\cite{ACE_Lin,ACE_Lin_1} operator formulation (r-RESPA+ACE).\cite{JCP_2019_sagar}
The implementation of our method in the CPMD code\cite{cpmd} allows us to perform long
hybrid functional based AIMD simulations.
The r-RESPA+ACE method has been shown to be efficient and accurate in predicting the structural and dynamical properties of bulk water.\cite{JCP_2019_sagar}

Here we use the r-RESPA+ACE operator scheme together with the well-sliced metadynamics  (WS-MTD)\cite{JCC_shalini} method to study a prototype hydrolysis reaction in water.
In particular, we model the formamide hydrolysis reaction in aqueous alkaline medium and compute the two-dimensional free energy surface for the reaction.
Finally, we compare the accuracy of PBE (GGA) and PBE0 (hybrid) functionals in
predicting the free energy surface.
\section{\label{sec:theory}Theory}
\subsection{Exact Exchange Operator}
The self consistent field (SCF) solution of hybrid functional based KS-DFT equations  
%
requires the application of the exchange operator on all the occupied KS orbitals at every SCF iteration. 
%
%
%
%
The exact exchange operator ${\mathbf V}_{\rm X}$ is defined as 
\begin{equation}
{\mathbf V}_{\rm X}= -\sum_{j}^{N_{\rm orb}} \frac{| \psi _{j} \rangle  \langle \psi _{j} |}{r_{12}} \enspace ,
\end{equation}
in terms of the set of occupied KS orbitals $\{|\psi_{j} \rangle\}$.
Here, $N_{\rm orb}$ is the total number of occupied orbitals and $r_{12}=\left | \mathbf r_1 - \mathbf r_2 \right | $.
The application of ${\mathbf V}_{\rm X}$ on a KS orbital $|\psi _{i} \rangle$ is given by,
\begin{equation}
\begin{split}
{\mathbf V}_{\rm X}|\psi _{i}\rangle & =- \sum_{j}^{N_{\rm orb}} |\psi _{j} \rangle \left \langle\psi _{j} \left | \left ( r_{12}\right )^{-1} \right | \psi _{i}\right \rangle 
\\ & =- \sum_{j}^{N_{\rm orb}} v_{ij}(\mathbf{r}_{1}) |\psi _{j}\rangle \enspace, ~~i=1,....,N_{\rm orb}
\end{split}
\end{equation}
where
\begin{equation}  
\label{e:vij}
v_{ij}(\mathbf {r}_{1})=\left \langle\psi _{j} \left | \left ( r_{12}\right )^{-1} \right | \psi _{i}\right \rangle \enspace.   
\end{equation}
%
%

%
%
%
%
%
%
%
%
%

%
%
%
%
%
The evaluation of $v_{ij}(\mathbf{r})$ is optimally done in the reciprocal space\cite{JCP_HFX_Voth,PRB_Car_Wannier} using Fourier transformation.
%
If $N_{\rm G}$ is the total number of PWs used, the computational cost for doing Fourier transform scales as $N_{\rm G}\log  N_{\rm G}$ using fast Fourier transform technique.
The application of ${\mathbf V}_{\rm X}$ on a single KS orbital requires $N_{\rm orb}$ times evaluation of $v_{ij}(\mathbf{r})$.
%
%
Thus, the total computational cost  scales as $N_{\rm orb}^2 N_{\rm G}\log  N_{\rm G}$,\cite{JCP_HFX_Voth} as ${\mathbf V}_{\rm X}$ has to be applied on $N_{\rm orb}$ number of KS orbitals.
For typical molecular systems of our interest, $N_{\rm orb} \sim 10^2$ and $N_{\rm G} \sim 10^6$,  which results in an exorbitant computational time requirement for hybrid functional calculations.
%

%

\subsection{Adaptively Compressed Exchange Operator}
%
%
%
Recently, Lin Lin developed the ACE operator formulation,\cite{ACE_Lin,ACE_Lin_1} to reduce the computational cost of such calculations.
%
In this formalism, ${\mathbf V}_{\rm X}$ operator is approximated by the ACE operator ${\mathbf V}_{\rm X}^{\rm ACE}$. 
%
%
First, the action of ${\mathbf V}_{\rm X}$ on the set of KS orbitals $\{|\psi_{i} \rangle\}$ is computed as 
\begin{equation}
|W_{i}\rangle={\mathbf V}_{\rm X}|\psi _{i}\rangle , \enspace ~~i=1,....,N_{\rm orb} \enspace . 
\end{equation}
According to the ACE formalism, ${\mathbf V}_{\rm X}^{\rm ACE}$ is defined as
\begin{equation}
{\mathbf V}_{\rm X}^{\rm ACE}= \sum_{i,j}^{N_{\rm orb}}  | W_{i} \rangle B_{ij}  \langle W_{j} | \enspace,
\end{equation}
where, ${\mathbf B}={\mathbf M}^{-1}$, 
%
and the elements of the matrix ${\mathbf M}$ are 
\begin{equation}
M_{kl}= \left \langle \psi_k | {\mathbf V}_{\rm X}|\psi _{l} \right \rangle   \enspace.
\end{equation}
Now, the Cholesky factorization of $-{\mathbf M}$ is performed as 
\begin{equation}
 {\mathbf M}=-{\mathbf L}{\mathbf L}^T  \enspace.
\end{equation}
Here, ${\mathbf L}$ is a lower triangular matrix.
Then ${\mathbf B}$ can be computed as
\begin{equation}
 {\mathbf B}=-{\mathbf L}^{-T}{\mathbf L}^{-1}  \enspace. 
\end{equation}
Now, the ${\mathbf V}_{\rm X}^{\rm ACE}$ operator can be rewritten as,
\begin{equation}
{\mathbf V}_{\rm X}^{\rm ACE}   = - \sum_{k}^{N_{\rm orb}} |P_{k} \rangle  \langle P_k |   \enspace.
\end{equation}
Here, the ACE projection vectors $\{|P_{k} \rangle\}$ are the columns of the matrix ${\mathbf P}$, which is defined 
as
\begin{equation}
  {\mathbf P}= {\mathbf W}{\mathbf L}^{-T} \enspace.
\end{equation}
Now, the evaluation of the action of ${\mathbf V}_{\rm X}^{\rm ACE}$ operator on KS orbitals can be performed with $N_{\rm orb}^{2}$ number of simpler inner products as
\begin{equation}
{\mathbf V}_{\rm X}^{\rm ACE}|\psi _{i}\rangle=- \sum_{k}^{N_{\rm orb}} |P_{k} \rangle  \left \langle P_k | \psi_{i} \right \rangle , \enspace ~i=1,....,N_{\rm orb}   \enspace.
\end{equation}
The advantage of the ACE approach is that the cost of applying the ${\mathbf V}_{\rm X}^{\rm ACE}$ operator on each KS orbitals is much less as compared to the application of the ${\mathbf V}_{\rm X}$ operator.
%

In the scheme proposed by Lin~Lin,\cite{ACE_Lin,ACE_Lin_1} at the beginning of the SCF iteration, ${\mathbf V}_{\rm X}^{\rm ACE}$ operator is constructed through the computation of $\{|W _{i}\rangle \}$ as described above. 
Certainly, this step is computationally costly due to $N_{\rm orb}^{2}$ times evaluation of $v_{ij}(\mathbf{r})$.
As the exchange energy has only a small contribution to the total energy, it is possible to use the same ${\mathbf V}_{\rm X}^{\rm ACE}$ operator (without updating it) for all the SCF iterations.
{An outer loop over the SCF calculation updates the ${\mathbf V}_{\rm X}^{\rm ACE}$ operator till a 
complete convergence is obtained.}
It has to be noted that, once the ${\mathbf V}_{\rm X}^{\rm ACE}$ operator is constructed, its low rank structure allows the easy computation of $\{{\mathbf V}_{\rm X}^{\rm ACE}|\psi _{i}\rangle \}$ during the SCF iterations.
{For e.g., for a 32 water periodic system, the computational time
for the operation of ${\mathbf V}_{\rm X}^{\rm ACE}$ is 
$\sim$240 times smaller than that of ${\mathbf V}_{\rm X}$ (using 120 compute cores).\cite{JCP_2019_sagar}}
%
%
%
%
%
%

\begin{figure*}
\includegraphics[scale=0.33]{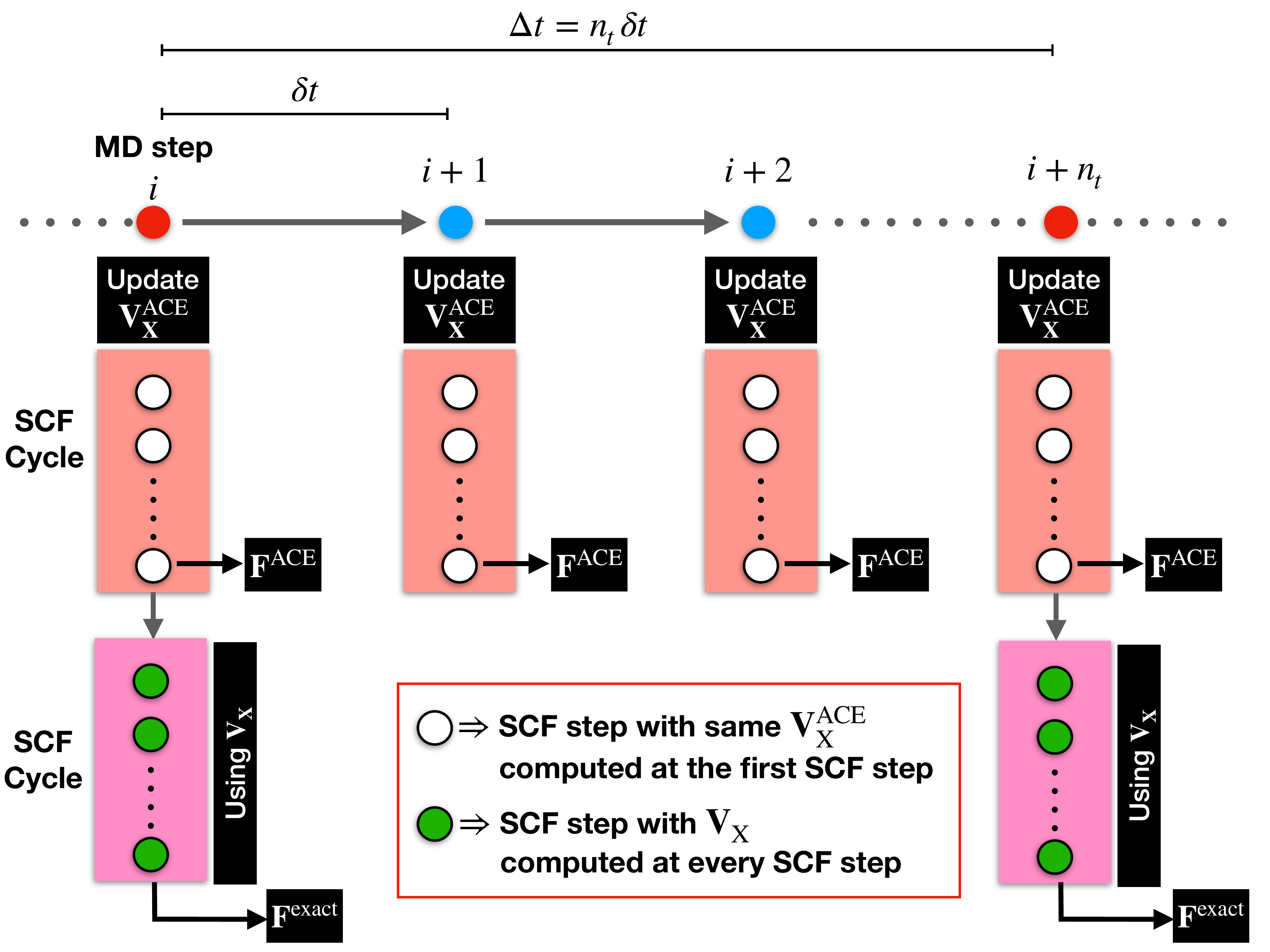}
\caption{\label{scheme} { The proposed r-RESPA+ACE scheme for performing AIMD simulations.}
}
\end{figure*}
\subsection{Combining ACE Operator Formalism with r-RESPA}
Recently, we took advantage of the properties of the ACE operator to combine it with r-RESPA method\cite{r-RESPA,Tuckerman-book} for reducing the computational cost of hybrid functional based AIMD calculations.\cite{JCP_2019_sagar} 
%
%
%
In that, we made the assumption that the ionic forces, $\{F_K\}$, with $K=1,\cdots,3N$, for a system containing $N$ particles, can be decomposed into slow and fast components depending on the time scale of their variation as 
\begin{equation}
F_K=F_K^{\rm fast}+F_K^{\rm slow} , \enspace ~~K=1,....,3N \enspace . 
\end{equation}
In this case, the Liouville operator can be written as,
\begin{equation}
\begin{split}
iL & =\sum_{K=1}^{3N}\left [ \dot{X}_K \frac{\partial}{\partial {X}_K} + F_{K}^{\rm fast} \frac{\partial}{\partial {P}_K} + F_{K}^{\rm slow} \frac{\partial}{\partial {P}_K} \right ]  \\ &
%
= iL^{\rm fast}_{1} + iL^{\rm fast}_{2} + iL^{\rm slow} \enspace,
\end{split}
\end{equation}
with
\begin{equation}
iL^{\rm fast}_{1} = \sum_{K=1}^{3N}\left [ \dot{X}_K \frac{\partial}{\partial {X}_K} \right ],~ iL^{\rm fast}_{2} = \sum_{K=1}^{3N} \left [ F_{K}^{\rm fast} \frac{\partial}{\partial {P}_K} \right ]
\end{equation}
and
\begin{equation}
iL^{\rm slow} = \sum_{K=1}^{3N}\left [ F_{K}^{\rm slow} \frac{\partial}{\partial {P}_K} \right ] \enspace.
\end{equation}
Here, $\{X_K\}$ and $\{P_K\}$ are the positions and the conjugate momenta of the particles, respectively.
%
%
%
%

The symmetric Trotter factorization of the classical propagator based on the above mentioned decomposition gives 
%
\begin{equation}
\begin{split}
%
\exp & (iL\Delta t)  \approx \exp \left (iL^{\rm slow} \frac{\Delta t}{2} \right) \\ &  \times \left [ \exp \left (iL^{\rm fast}_2 \frac{\delta t}{2} \right)   \exp \left(iL^{\rm fast}_1\delta t \right)  \exp \left(iL^{\rm fast}_2 \frac{\delta t}{2} \right)  \right ]^{n_t} \\ & \times  \exp \left (iL^{\rm slow} \frac{\Delta t}{2} \right) \enspace.
%
\end{split}
\end{equation}
%
%
%
Here, $n_t$ is a natural number, while the choice of larger time step $\Delta t$ is decided based on the time scale of the variation of slow forces ($\{F_K^{\rm slow}\}$) and 
the choice of smaller time step $\delta t=\Delta t/n_t$ is decided based on the time scale of fast forces ($\{F_K^{\rm fast}\}$).
%
%
%
%

%
%
%
%
As next, we propose an approximate splitting  of ionic forces,\cite{JCP_2019_sagar} as 
%
%
\begin{equation}
    F^{\rm exact}_K= F^{\rm ACE}_K+ \Delta F_K  \enspace , \enspace K=1,\cdots,3N \enspace 
\end{equation}
with $\Delta F_K = \left ( F^{\rm exact}_K -  F^{\rm ACE}_K \right )$,  
${\mathbf F}^{\rm exact}$ is the ionic force computed using the full rank exchange operator ${\mathbf V}_{\rm X}$ and $\mathbf F^{\rm ACE}$ is the ionic force computed using
the low rank ${\mathbf V}_{\rm X}^{\rm ACE}$ operator. 
Further, we consider,
\begin{eqnarray}
F^{\rm slow}_K &\equiv& \Delta F_K , \mathrm{and} \nonumber \\[1ex]
F^{\rm fast}_K &\equiv& F^{\rm ACE}_K \enspace .
\end{eqnarray}

{ In our implementation, as shown in FIG.~\ref{scheme}, the ${\mathbf V}_{\rm X}^{\rm ACE}$ operator is constructed at the beginning of 
every SCF cycle, based on the initial guess of the wavefunction.
The same ${\mathbf V}_{\rm X}^{\rm ACE}$ operator is used
in the remaining SCF steps, till a convergence in wavefunction is achieved.
The initial wavefunction for the SCF cycles is obtained from the 
Always Stable Predictor Corrector Extrapolation scheme\cite{JCC_ASPC}.
The converged wavefunction is then used to compute $\mathbf F^{\rm ACE}$.
However, for every $n_t$ MD steps, another SCF cycle is executed
where the exact exchange operator $\mathbf V_{\rm X}$ is computed
at every SCF step.
The converged wavefunction is then used to compute $\mathbf F^{\rm exact}$, and then $\Delta \mathbf F$.}

%
%
%
%
As the ${\mathbf V}_{\rm X}^{\rm ACE}$ operator closely resembles the ${\mathbf V}_{\rm X}$ operator,
 the differences in the ionic force components of $\mathbf F^{\rm exact}$ and $\mathbf F^{\rm ACE}$ are very small.
%
In our earlier work,\cite{JCP_2019_sagar} we demonstrated that the magnitude of $\Delta \mathbf F$ is $\sim$100 times smaller than that of
$\mathbf F^{\rm ACE}$. 
It was also shown that $\Delta \mathbf F$ computed at every $n_t = \Delta t / \delta t$ steps is slowly varying as compared to $\mathbf F^{\rm ACE}$.
%

%
%
%
%
%
%
%

\begin{figure}[t]
\includegraphics[width=0.45\textwidth]{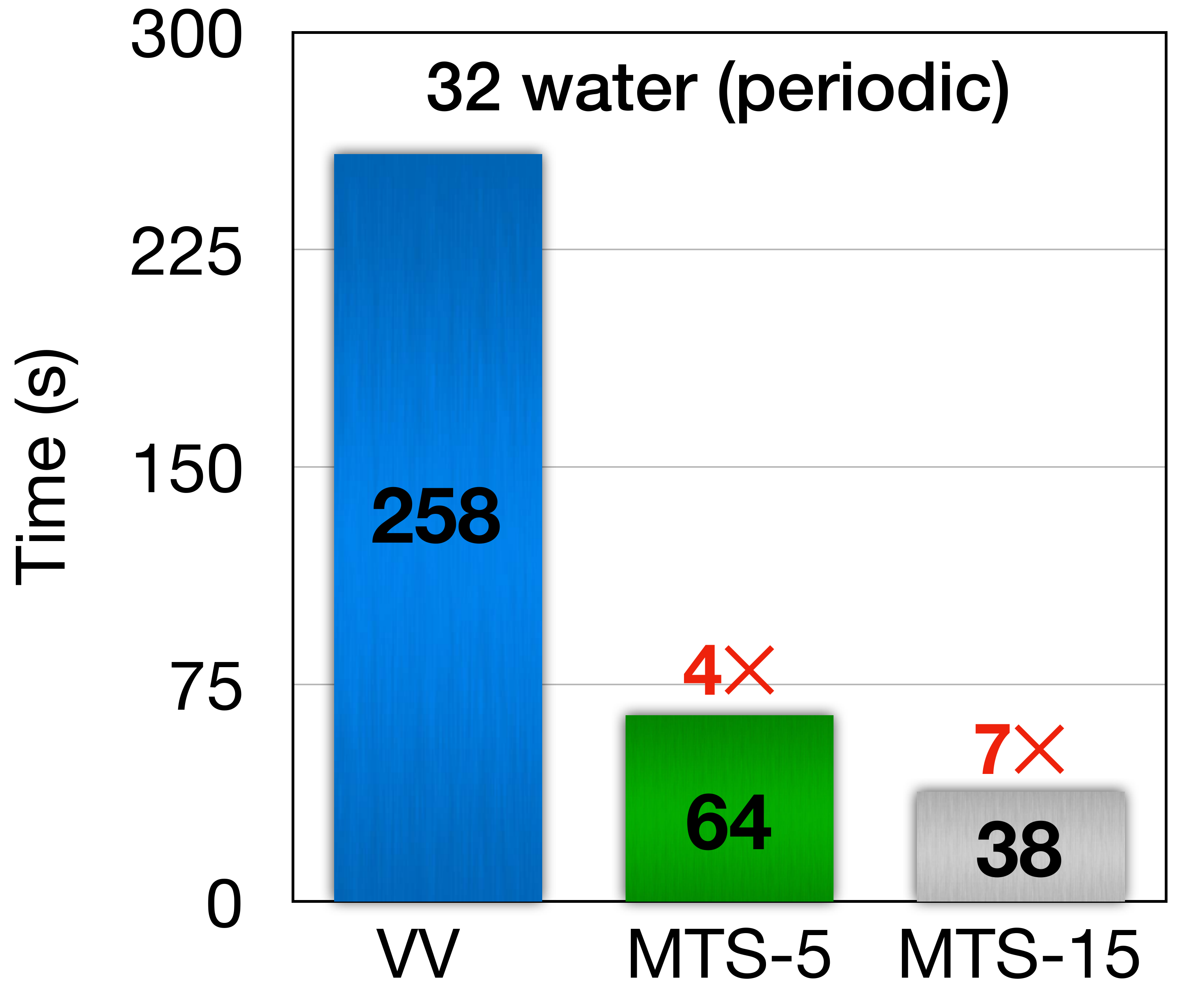}%
\caption{\label{timings} 
Average computational time per MD step for PBE0 calculation
of a system containing 32 water molecules in a $10\times 10 \times 10$~{\AA}$^3$ periodic box
using various methods:
Using the exact $\mathbf V_{\rm X}$ operator ({\bf VV}); using r-RESPA+ACE with $n_t = 5$ ({\bf MTS-5}), and $n_t=15$ ({\bf MTS-15}).
All computations were using identical 120 processors.
The computational time reported here is averaged over 500 MD steps.
The achieved speed-ups for {\bf MTS-5} and {\bf MTS-15}, compared to {\bf VV}, are shown in red.   
This data is taken from Ref.~\onlinecite{JCP_sagar}.
}
\end{figure}

%
%
%
In this way, r-RESPA+ACE scheme allows us to compute computationally costly $\Delta \mathbf F$ (or $\mathbf F^{\rm exact}$) less frequently than the computationally cheaper $\mathbf F^{\rm ACE}$. 
As a result,   
we get a substantial speed-up.
For a periodic system containing 32 water molecules, we obtained a speed-up of 7 with $n_t=15$; See FIG.~\ref{timings} and Ref.~\onlinecite{JCP_2019_sagar}.
%
%
%
%
Larger values of $n_t$ could be used within our scheme by 
eliminating resonance effects with the aid of  thermostats.\cite{mts_resonance,Resonance_MET} 


\section{Computational Details}
%
For modelling the chemical reaction, a cubic periodic simulation cell with a side length of 10~{\AA} was chosen, which contained one formamide molecule, one hydroxide ion, and 29 water molecules.
All the calculations were carried out employing
the {\tt CPMD} program\cite{cpmd} wherein the r-RESPA+ACE method was  implemented by us.
%
The PBE0\cite{JCP_PBE0_model} XC functional was used together with the norm-conserving Troullier-Martin type pseudopotentials\cite{PRB_TM}
and a PW cutoff energy of 80~Ry was taken.
%
We carried out Born-Oppenheimer molecular dynamics simulations to perform MD simulations at canonical (NVT) ensemble for $T=300$K.
Here, r-RESPA+ACE scheme with $\delta t = 0.48$~fs and  $\Delta t = 7.2$~fs (i.e. $n_t=15$) was considered and Nos{\'e}--Hoover chain thermostats were used.\cite{NHC} 
%
During the SCF (see FIG.~\ref{scheme}), we converged the wavefunctions till the magnitude of the maximum wavefunction gradient 
was below $1\times 10^{-6}$ au.
%
Always Stable Predictor Corrector Extrapolation scheme\cite{JCC_ASPC} of order 5 was used to obtain initial guess of wavefunction.

We employed the WS-MTD approach\cite{JCC_shalini} to compute the free energy surface of the base-catalyzed formamide hydrolysis reaction.
The WS-MTD method is designed to achieve a controlled sampling of coordinates and efficient exploration of high-dimensional
free energy landscape.
This method is ideal for investigating the formamide hydrolysis reaction, 
since we would like to sample the distance between a reactive water molecule and the formamide in a controlled manner, while ensuring exhaustive sampling of 
the protonation state of the attacking water.
We chose two collective variables (CVs), ${\mathbf s =  \{ s_1, s_2 \} }$, to explore the free energy surface of the reaction.
%
%
The first CV ($s_1$) is the distance, $d$[C--O$_1$], between the carbon atom (C) of the formamide and the oxygen atom (O$_1$) of the attacking water molecule, as shown in FIG.~\ref{mech}.
This CV was sampled using the umbrella sampling\cite{US_method} like 
bias potential
%
%
\begin{eqnarray} 
\label{e:us:bias}
W_h(s_1)=\frac{1}{2}\kappa_h \left ( s_1 - d_h^{(0)} \right )^2 , ~~h=1,...,M \enspace 
\end{eqnarray}
which ensures a controlled sampling along this coordinate.
Here, $M$ is the total number of umbrella windows used, while
$\kappa_h$ and $d^{(0)}_h$ are the restraining force constant and the equilibrium value of the $h$-th umbrella restraint, respectively.
The second CV ($s_2$) is the coordination number ($CN$) of the oxygen atom (O$_1$) of the attacking water with all the hydrogen atoms (H$_{\textrm w}$) of the solvent molecules (including that of itself): 
\begin{equation}
\label{CN}
CN[{\mathrm {O_1:H_w}}]=\sum_{i=1}^{N_{\mathrm {H_w}}} \frac{ 1}{1+({d_{1i}}/{d_0})^{6}} \enspace ,
\end{equation}
with $d_0=1.30$~{\AA}.
Here, $N_{\mathrm {H_w}}$ is the total number of H$_{\textrm w}$ atoms and $d_{1i}$ is the distance between the O$_1$ atom and the $i$-th H$_{\textrm w}$ atom.
%
Well tempered metadynamics (WT-MTD) bias potential\cite{WT-MTD}, $V^{\rm b}$, was employed to sample $s_2$:
%
\begin{eqnarray} 
\label{e:mtd:bias}
V^{\rm b}(s_2,t) = \sum_{\tau < t} w(\tau) \exp \left [ -\frac{ \left \{ s_2 - s_2(\tau) \right \}^2 }{2 \left ( \delta s \right )^2 } \right ] \enspace. \end{eqnarray}
Here, $\delta s$ is the width of the Gaussian function and the height of the Gaussian 
$w(\tau)$ is given by,
\begin{equation}
 w(\tau) = w_0 \exp \left [ - \frac{ V^{\rm b}(s_2,\tau) }{k_{\rm B} \Delta T} \right ] \enspace,
 \end{equation}
where $w_0$ and $\Delta T$ are parameters and $k_{\rm B}$ is the Boltzmann constant.

The WS-MTD approach allows us to perform $M$ number of independent simulations for each umbrella in parallel. 
AIMD Lagrangian of an umbrella window $h$ is given by,
\begin{eqnarray}
 \mathcal L^{\rm WS-MTD}_h = \mathcal L^{(0)} - W_h(s_1) - V^{\rm b}(s_2,t)
\end{eqnarray}
where $\mathcal L^{(0)}$ is the original AIMD Lagrangian  
while 
$W_h(s_1)$  and $V^{\rm b}(s_2,t)$ are the biases, as discussed earlier.
We obtain the biased probability distributions $\{ \tilde P_h(s_1,s_2) \}$ from these simulations, 
which is then reweighted and combined using the Weighted Histogram Analysis Method,\cite{WHAM_JCC} as discussed in Ref.~\onlinecite{JCC_shalini}. 
Using the 
reweighted distribution $P(s_1,s_2)$ thus obtained, 
%
%
the two-dimensional 
free energy surface is constructed as
\[ F(s_1,s_2) = - k_{\rm B} T \ln P(s_1,s_2) \enspace . \]

In total, 29 umbrella windows were used for sampling $d$[C--O$_1$] in the range from 1.51 to 3.70~{\AA}.
The parameters of the umbrella bias potentials, $\kappa_h$ and $d_h^{(0)}$, were taken according to our earlier study.\cite{JCP_sagar}  
%
The time-dependent WT-MTD bias acting along $CN[{\mathrm {O_1:H_w}}]$
was updated every 19.4~fs.
%
The WT-MTD bias parameters $w_0$, $\delta s$ and $\Delta T$ were chosen to be 0.59~kcal~mol$^{-1}$, 0.05 and 4000~K, respectively. 
After carrying out 2--3~ps of equilibration for each umbrella window,
%
we performed 10~ps of production run for every window. 
Thus a total of $29\times 10$~ps long production trajectory
was obtained at the level of hybrid-DFT.
The initial structure for an umbrella window was taken from the equilibrated structure of the adjacent window.
%

%
\section{\label{sec:result_diss}Results and Discussion}

%
%
\begin{figure}
\includegraphics[scale=0.65]{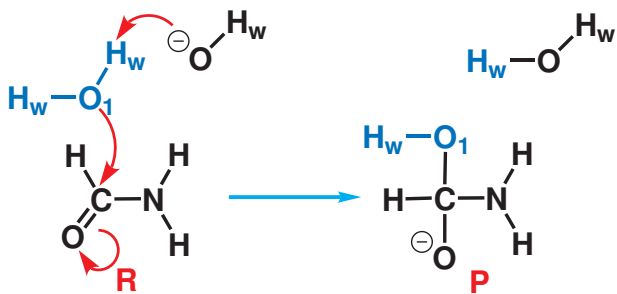}
\caption{\label{mech} The mechanism of formation of the tetrahedral intermediate {\bf P} from the reactant {\bf R}
during the formamide hydrolysis in 
aqueous alkaline medium.
%
%
}
\end{figure}
%
%
%
%
%
%
\begin{figure}
\includegraphics[scale=0.65]{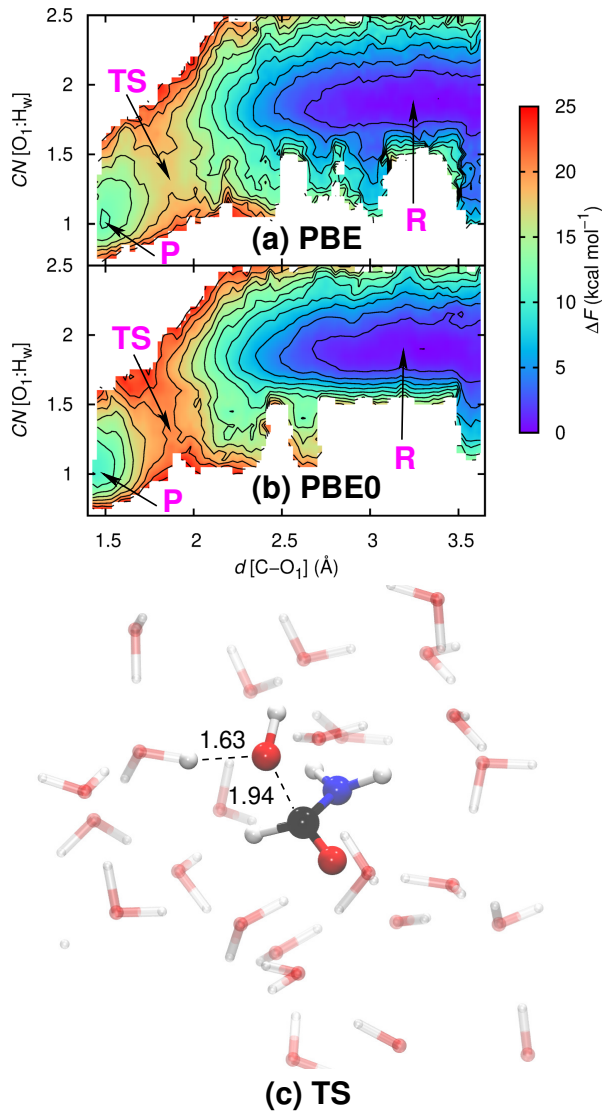}
\caption{\label{fes} Free energy surfaces computed using 
(a) {PBE},\cite{JCP_sagar}
and 
(b) { PBE0}
density functionals for the {\bf R}$\rightarrow${\bf P} reaction are presented. Contour lines are drawn every 2 kcal mol$^{-1}$. A representative snapshot of the {\bf TS} state
is shown in (c). The indicated bond distances here are in {\AA}. Atom colors: red (O); blue (N); black (C); white (H).}
\end{figure}
\begin{figure}
\includegraphics[scale=0.65]{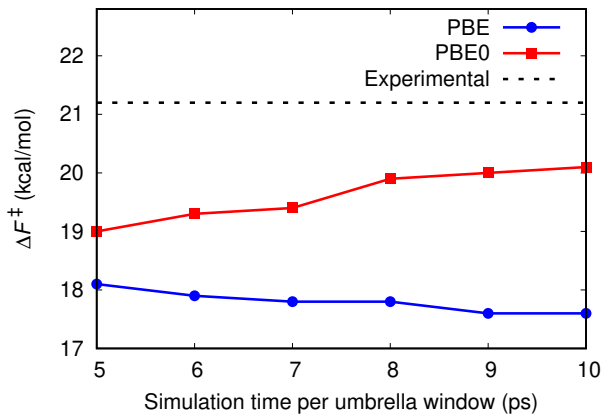}
\caption{\label{fes_conv} Convergence of the free energy barrier ($\Delta F^\ddagger$) {\bf R}$\rightarrow${\bf P} reaction with simulation time (per umbrella window) using {PBE},\cite{JCP_sagar} and {PBE0} density functionals. The experimental measure of $\Delta F^\ddagger$ is shown in dotted black line.
}
\end{figure}
\begin{table}
\caption{\label{table} Free energy barriers ($\Delta F^\ddagger$) of 
the reaction {\bf R}$\rightarrow${\bf P} (FIG.~\ref{mech}) 
using PBE and PBE0 functionals are compared
with the experimental measure.}
\begin{ruledtabular}
\begin{tabular}{lcr}
Method & $\Delta F^\ddagger$ (kcal mol$^{-1}$)\\
\hline
{ PBE}\cite{JCP_sagar} & 17.6  \\
{ PBE0} & 20.1  \\
Experiment\cite{Formamide_Exp} & 21.2{$\pm$0.2}  \\
\end{tabular}
\end{ruledtabular}
\end{table}
The alkaline hydrolysis of formamide in aqueous medium is one of the basic reactions that is of interest in the field of chemistry and biochemistry.
In particular, it serves as a model for the hydrolysis of peptide bonds in enzymes.
The reaction has been well studied experimentally\cite{Formamide_Exp} and theoretically,\cite{JPCB_Carloni,CPL_Klein,Angew_Klein,Pietrucci15030} and it is an ideal prototype reaction in liquid water to demonstrate the application of our method.
%

We launched WS-MTD simulations to investigate this reaction.
%
The hydrolysis reaction proceeds through the nucleophilic attack of the hydroxide ion on the carbon atom of the carbonyl group to form a tetrahedral intermediate ({\bf P}) as shown in FIG.~\ref{mech}.
%
The rate limiting step of the hydrolysis is the formation of the tetrahedral intermediate ({\bf P}).\cite{CPL_Klein,Angew_Klein} 
%

The computed free energy surface is given in FIG.~\ref{fes}. 
The minimum energy pathway on this surface was used to extract the mechanism and free energy barrier.
The convergence of the free energy barriers with simulation time is shown in 
FIG.~\ref{fes_conv} and the converged free energy barriers are listed in Table~\ref{table}.
For comparing with the PBE data, we have taken the results
from our earlier work.\cite{JCP_sagar}

In both PBE and PBE0 free energy surfaces, the deep minimum corresponding to the reactant state {\bf R} is nearly at the same location: $s_1 \in [2.9,3.5]$({\AA}), $s_2 \in [1.8,2.0]$ (unitless).
The state {\bf P} is the tetrahedral intermediate, and the locations of {\bf P} are also nearly the same for both the functionals: $s_1 \sim 1.5$({\AA}), $s_2 \sim 1.0$ (unitless). 
Similarly, the saddle point on the landscape ({\bf TS}) is also  nearly at the same locations on the two surfaces: $s_1 \sim 1.9$({\AA}), $s_2 \sim 1.3$ (unitless). 
 However, the passage through the {\bf TS} state is narrower on the {PBE0} surface than on the {PBE} surface. 
As expected, in the {\bf TS} structure, the attacking water molecule has dissociated one of its proton, 
and a weak covalent bond between O$_1$ and the carbon atom is formed  (see FIG.~\ref{fes}(c)).
This mechanism is in agreement with previous computations.\cite{Angew_Klein}

Although the topology of the free energy surfaces are qualitatively similar for the 
two functionals, the free energy barrier for the reaction differs substantially between the two.
%
%
%
%
The free energy barrier for {PBE} is 17.6~kcal/mol, while that for {\bf PBE0}  
is 20.1~kcal/mol.  
The PBE0 free energy barrier is 2.5~kcal/mol higher than that of PBE.
Most importantly, the barrier computed from PBE0 is closer to the experimentally 
determined free energy barrier of 21.2{$\pm$0.2}~kcal/mol at  $298$~K.\cite{Formamide_Exp}
For the same reaction, similar trend in PBE and PBE0 free energies was noticed in our earlier work,\cite{JCP_sagar} in which the PBE0 calculations were carried out using a  
Noise Stabilized Molecular Dynamics approach and localized KS orbitals.
%
%
%

\section{\label{sec:concl}Conclusions}
By combining r-RESPA and ACE formalism, an effective speed-up of
7 was obtained in AIMD simulations using hybrid functionals and plane waves.
With such a speed-up, we are able to perform free energy calculations for chemical reactions in water.
We demonstrated this by studying the base catalyzed
hydrolysis of formamide in water.
Although the system contained $\sim 100$ atoms, 
we could perform $29 \times 10$~ps 
long WS-MTD simulations with PBE0 functional.
%
Based on these calculations, the two-dimensional
free energy surface for the chemical reaction was 
computed.
We find that PBE (GGA) functional underestimates the 
free energy barrier by $\sim$3~kcal/mol.
On the other hand, PBE0 (hybrid) functional is
able to predict the free energy barrier more accurately - it is only $\sim$1~kcal/mol lower than the experimental measure of the free energy barrier.
In conclusion, we have demonstrated that free energy computations of chemical reactions
are now computationally affordable 
with hybrid functionals and plane wave basis set employing the r-RESPA+ACE approach.


\begin{acknowledgments}
Authors acknowledge the HPC facility at the Indian Institute of Technology Kanpur (IITK) for the computational resources.
%
SM thanks the University Grant Commission (UGC), India, for his Ph.D. fellowship.
\end{acknowledgments}



%
\bibliography{aipsamp}

\end{document}